\documentclass[12pt]{article}

\usepackage{scicite}
\usepackage{multicol}
\usepackage{graphicx}
\usepackage{times}
\usepackage{geometry}
\usepackage{caption}
\usepackage{booktabs}
\usepackage{hyperref}
\usepackage{subcaption}
\usepackage[font=small,labelfont=bf]{caption}

\usepackage[scaled=0.95]{inconsolata}
\newenvironment{sciabstract}{%
\begin{quote} \bf}
{\end{quote}}

\newcounter{lastnote}

\title{Evidence of economic segregation from mobility lockdown during COVID-19 epidemic}

\author
{Giovanni Bonaccorsi$^{1\ast}$, Francesco Pierri$^{2}$, Matteo Cinelli$^{3}$, Francesco Porcelli$^{4}$, \\
Alessandro Galeazzi$^{5}$, Andrea Flori$^{1}$, Ana Lucia Schmidt$^{6}$, Carlo Michele Valensise $^{7}$, \\
Antonio Scala$^{3}$, Walter Quattrociocchi$^{6\ast}$, Fabio Pammolli $^{1,8\ast}$\\
\\
\normalsize{$^{1}$ Impact, Department of Management, Economics and Industrial Engineering, Politecnico di Milano}\\
\normalsize{$^{2}$ Department of Electronics, Information and Bioengineering, Politecnico di milano}\\
\normalsize{$^{3}$ CNR-ISC}\\
\normalsize{$^{4}$ Department of Law, Università di Bari }\\
\normalsize{$^{5}$ Department of Information Engineering, Università di Brescia, }\\
\normalsize{$^{6}$ Department of Environmental Sciences, Informatics and Statistics, Università Ca'Foscari di Venezia}\\
\normalsize{$^{7}$ Department of Physics, Politecnico di Milano}\\
\normalsize{$^{8}$ CADS, Joint Center for Analysis, Decisions and Society, Human Technopole,Politecnico di Milano} 
\\
\normalsize{$^\ast$To whom correspondence should be addressed; E-mail:  .}
}

\date{}

\begin{document} 


\baselineskip24pt


\maketitle


\begin{sciabstract}
\normalsize
In response to the COVID-19 pandemic, National governments have applied lockdown restrictions to reduce the infection rate.
We perform a massive analysis on near real-time Italian data provided by Facebook to investigate how lockdown strategies affect economic conditions of individuals and local governments.  We model the change in mobility as an exogenous shock similar to a natural disaster. We identify two ways through which mobility restrictions affect Italian citizens. First, we find that the impact of lockdown is stronger in municipalities with higher fiscal capacity. Second, we find a segregation effect, since mobility restrictions are stronger in municipalities for which inequality is higher and where individuals have lower income per capita. 

 
\end{sciabstract}

\section*{Introduction}
On March 9th 2020, Italy was the first European country to apply a national lockdown \cite{dpcm9marzo} in response to the spread of novel coronavirus (COVID-19) outside China borders.
%
%
Following Italy and China, national lockdowns have been adopted by other governments, and mobility flows have been drastically reduced to lessen the reproduction rate of COVID-19\cite{google}.

Increasing concerns are arising on the economic consequences of lockdown and how it can disproportionally affect the weaker and the poorer \cite{nytimes}. Policy restrictions imposed by lockdown measures have determined a detrimental effect on several production sectors, heavily deteriorating global value chains and trade exchanges, motivating governments to announce fiscal interventions of about \$8 trillion and massive monetary measures from the G20 and others \cite{imfcovid}. Supply shocks can, in fact, trigger variations in aggregate demand that ultimately can be even larger than the COVID-related shocks themselves, hence imposing to immediately incorporating principles of system resilience to systemic disruption in order not to condition the future socio-economic recovery for the next decade \cite{anderson2020will,atkeson2020will,guerrieri2020macroeconomic,williamnew}. 

The intensity of the sudden stop induced by the Covid-19 outbreak produces effects which are similar to those produced by a large scale natural disaster \cite{chiaromontepnas, Carvalho_Nirei_Saito_Tahbaz-Salehi_2016, Boehm_Flaaen_Pandalai-Nayar_2018, Inoue_Todo_2019}. Here, analogously to the literature on the economic damages of natural disasters \cite{noy2009macroeconomic,kellenberg2008does, chiaromontepnas}, we model the change in mobility affecting Italian municipalities as an exogenous shock.  
We leverage a de-identified large-scale collection of near real-time data provided by Facebook platform to characterize the effect of population mobility restrictions \cite{maas2019facebook}. On economic data, we rely on official statistics at highest available level of resolution, i.e. municipalities, to investigate the features of those who are mostly affected. 

To understand how the lockdown measures impact on the economy, we rely as a proxy for economic downturn on mobility variations between and within italian municipalities across the deployment of the lockdown announcement. As shown in \cite{google}, mobility trends have reduced in fact by more than 90\% in Italy after the lockdown, both in the retail and tourism sectors and in the service one, and this disruption of the mobility toward the workplace supports the use of mobility flows as a proxy of economic damages. Furthermore, we want to investigate the geographic distribution of the shocks in order to identify the economic conditions of the most and least affected zones. Rather than a homogeneous distribution we find, on the one side, that mobility reduction induced by lockdown is stronger for municipalities with higher fiscal capacity. On another side, we find that the contraction in mobility is higher for municipalities with lower per capita income and for those with higher inequality. 
In the aftermath of the crisis, central governments need not only to sustain economic recovery, but also to compensate the loss of local fiscal capacity, while channeling resources to mitigate the impact of lockdown on poverty and inequality.


\section*{Results and Discussion}

\subsection*{Mobility restrictions}
In Figs. 1A-B we compare two daily snapshots of the mobility network of municipalities aggregated at province level. After 21 days of national lockdown, we notice a striking fragmentation of the usual national mobility pattern from North to South. 
We characterize daily connectivity patterns \cite{sciencemobility} through network measures \cite{barabasi2016network,newman2018networks}. 

In Fig.~1C, we analyze the temporal evolution of the number of the weakly connected components and the size of the largest connected component in the overall mobility network. We identify two opposite and significant trends, respectively increasing (Mann-Kendall (M-K): $P\sim0$; Kendall's Tau (K-T): $P\sim 0$ $R=0.64$; Theil-Sen (T-S): $R=30.52$) and decreasing (M-K: $P\sim0$; K-T: $P\sim 0$, $R=-0.67$; T-S: $R=-58.58$), that confirm the breakdown of hubs and long-range connections. 
We further assess the impact of mobility restrictions leveraging a network-based representation of mobility data and computing the network efficiency (Materials and Methods).
Efficiency~\cite{latora2001efficient} is a good proxy of the system dynamics at play. Indeed, it is a global network measure that combines the information deriving from the network cohesiveness and the distance among the nodes and it measures how efficiently information/individuals may travel over the network~\cite{crucitti2006centrality}. Additionally, it is particularly suitable for treating graphs with multiple components that evolve over time~\cite{bullmore2009complex}. As shown in Fig. 2, we observe a drastically decreasing trend of the efficiency (M-K: $P\sim0$; K-T: $P\sim 0$, $R=-0.75$; T-S: $R=-0.00003$) that confirms a pronounced drop in the mobility potential of the network.
%
Finally, we observe significant changes in the distribution of several node centrality measures over time (Materials and Methods, Figs.S1 to S9), with the most peripheral municipalities being those most affected by the lockdown. 

In the following, we use the variation in Nodal Efficiency (Material and Methods), that is the contribution of each node to the global network efficiency, 
as a proxy for the effects of mobility restrictions. We compute the percentage relative change induced by the lockdown, by constructing mobility networks in two windows, 15 days before and after the day of intervention.

\subsection*{Mobility lockdown and economic segregation}
\subsubsection*{Economic indicators}

First, we focus on average individual Income, which constitute the base for the Italian Personal Income Tax declared annually by taxpayers, as a measure of private resources available to individuals. 

Second, we use a municipal composite index of material and social well-being  (Index of Socio-economic Deprivation) produced by the Italian Ministry of Economy and Finance, which aggregates several dimensions of lack of material and social well-being at municipal level \cite{deprivazione,caranci_[italian_2010} (see Materials and Methods), representing one of the determinants of municipal standard expenditure needs. 

In Fig.~3 we show the relationships between the two indexes with respect to the relative change in Nodal Efficiency. We observe a significant correlation, negative with the Deprivation Index (Pearson: $R=-0.153$, $P\sim0$; S-R: $R=-0.235$, $P\sim0$; T-S: $R=-0.064$; K-T: $R=-0.162$, $P\sim0$) and positive with Income per Capita (Pearson: $R=0.263$, $P\sim0$; S-R: $R=0.404$, $P\sim0$; T-S: $R=0.001$; K-T: $R=0.273$, $P\sim0$). We also notice similar and significant relationships using other network centrality measures (Table S3).

As a third main variable, we consider the level of municipal Fiscal Capacity, measured each year by the Italian Ministry of Economy and Finance and employed in the fiscal equalization process. Municipalities with high Fiscal Capacity tend to be financially independent from central government to fund local expenditures. 

Finally, as additional regressors, we use a measure of municipal Inequality, i.e. the ratio between mean and median individual income, and an inverse measure of Urban Density in terms of the number of Real Estates per Capita.



\subsubsection*{Economic segregation from mobility disruption}

The joint distribution by percentiles of the variation of mobility and economic indicators is concentrated on the top and bottom percentiles (Figures S12-S13). This result is important, since it shows a different relation between the extremes of the distribution of economic indicators and mobility with respect to the one observed around the mean (Fig.~\ref{correlation}). 


Against that background, in Table \ref{qreg_m2} we show the results of a quantile regression, where the relative variation in Nodal Efficiency over time is regressed against a set of economic regressors with regional controls. The quantile regression approach relaxes the assumptions of linear regressions and estimates the conditional quantile of a dependent variable over its predictors instead of conditioning on the mean \cite{Koenker_Hallock_2001}. This allows us to concentrate on the dynamic at the tails of the distribution and to capture effects that with linear methods would otherwise seem insignificant \cite{chiaromontepnas}. Indeed our estimates at the top and bottom quantile of the distribution of the variation in Nodal Efficiency show a better fitting with respect to the OLS ones reported as reference in Table \ref{qreg_m2}


We observe a significant and positive relation between change in mobility during the lockdown and average individual Income for the bottom quantiles of the distribution. 
We study how municipalities at the lower end of the distribution of changes in mobility (10th-20th percentiles) are distributed according to their Income per Capita, and we find that the reduction in connectivity and mobility is higher for municipalities with a low average individual Income, while municipalities with high Income per Capita experience less intense changes. Moreover, at the upper end of the distribution, the relation is reversed. This asymmetry of the joint distribution of mobility contraction and Income per Capita unravels the existence of a possible segregation effect: even though some of the richer urban centers have experienced greater casualty rates, low income individuals are more affected by the economic consequences of the lockdown.




When we move to the analysis of municipal characteristics, measured through Deprivation and Fiscal Capacity, we find a different result: municipalities relatively richer in terms of social indicators and availability of own financial resources are those more hit by the loss in mobility efficiency in the aftermath of the lockdown. In other words, municipalities which experience the strongest effect of the lockdown are those with higher Fiscal Capacity and lower aggregate Deprivation.


Two seemingly opposite patterns emerge: individual indicators (average Income) show that the poorest are those more exposed to the economic consequences of the lockdown; on the contrary, aggregate indicators at the level of municipalities, i.e. Deprivation and Fiscal Capacity, reveal that richer municipality are those more hit by mobility contraction induced by the lockdown.

In order to shed light on these apparently contrasting results we look at the relationship between Inequality and the mobility contraction: we find a significant and negative relationship between the two variables at the lower end of the distribution of mobility reduction (10th-20th percentiles).

This result further characterizes our findings: not only stronger changes in mobility are associated with low income municipalities but they are also linked with high level of inequality. This happens together with a lower level of the Deprivation Index and the high level of Fiscal Capacity and introduce an important additional detail: the distribution on income. By controlling for all these factors, we finally see that municipalities most severely affected are those where economic distances among individuals are still significant and hence where an erosion of the supply of public services will have a greater impact.   

Finally, we find a negative relation between the number of building per capita and changes in mobility: municipalities affected more by the contraption in mobility have more buildings per capita, hence less Urban Density. 

Our evidence shows that the lockdown seems to produce an asymmetric impact, hitting poor individuals within municipalities with strong fiscal capacity. 

\section*{Conclusions}

We analyze Italian mobility data before and after the lockdown introduced to face the COVID-19 pandemic. We highlight how variations in mobility relate to some fundamental economic variables and we show that reduction in connectivity tend to be stronger for municipalities with low average individual Income and high income Inequality. At the same time, we show that the reduction in connectivity tends to be higher for municipalities with higher Fiscal Capacity. 

Our findings start to shed light on some social and economic consequences of policy measures adopted to contain the diffusion of COVID-19. 
First, the lockdown seems to unevenly affect the poorer fraction of the population. On another side, we find that the reduction in mobility and connectivity induced by the lockdown is more pronounced for municipalities with stronger Fiscal Capacity. Finally, the distribution of income plays a role: municipalities where inequality is greater have experienced stronger change in mobility and their citizens are more at risk.

The policy implications of our results suggest the necessity of asymmetric fiscal measures. Emergency grants should be channeled to poor people to support their consumption and, at the same time, to rich municipality to compensate the loss of fiscal capacity. In the absence of targeted lines of intervention, the lockdown would induce a further increase in poverty and inequality. 


\bibliography{scibib}

\begin{thebibliography}{10}

\bibitem{dpcm9marzo}
{Decreto del Presidente del Consiglio dei Ministri}, {\it Gazzetta Ufficiale\/}
  {\bf 62} (2020).

\bibitem{google}
Google, Covid-19 community mobility report,
  \url{https://www.gstatic.com/covid19/mobility/2020-03-29_IT_Mobility_Report_en.pdf}
  (2020).

\bibitem{nytimes}
D.~L. Jennifer Valentino-DeVries, G.~J. Dance, Location data says it all:
  Staying at home during coronavirus is a luxury,
  \url{https://www.nytimes.com/interactive/2020/04/03/us/coronavirus-stay-home-rich-poor.html}
  (2020).

\bibitem{imfcovid}
K.~Georgieva, {\it International Monetary Fund, Speech April 9th\/}  (2020).

\bibitem{anderson2020will}
R.~M. Anderson, H.~Heesterbeek, D.~Klinkenberg, T.~D. Hollingsworth, {\it The
  Lancet\/} {\bf 395}, 931 (2020).

\bibitem{atkeson2020will}
A.~Atkeson, What will be the economic impact of covid-19 in the us? rough
  estimates of disease scenarios, {\it Tech. rep.\/}, National Bureau of
  Economic Research (2020).

\bibitem{guerrieri2020macroeconomic}
V.~Guerrieri, G.~Lorenzoni, L.~Straub, I.~Werning, Macroeconomic implications
  of covid-19: Can negative supply shocks cause demand shortages?, {\it Tech.
  rep.\/}, National Bureau of Economic Research (2020).

\bibitem{williamnew}
OECD, {\it New Approaches to Economic Challenges (NAEC)\/}  (2020).

\bibitem{chiaromontepnas}
M.~Coronese, F.~Lamperti, K.~Keller, F.~Chiaromonte, A.~Roventini, {\it
  Proceedings of the National Academy of Sciences\/} {\bf 116}, 21450 (2019).

\bibitem{Carvalho_Nirei_Saito_Tahbaz-Salehi_2016}
V.~M. Carvalho, M.~Nirei, Y.~U. Saito, A.~Tahbaz-Salehi, {\it Supply Chain
  Disruptions: Evidence from the Great East Japan Earthquake\/}, no. ron287 in
  Discussion papers (2016).

\bibitem{Boehm_Flaaen_Pandalai-Nayar_2018}
C.~E. Boehm, A.~Flaaen, N.~Pandalai-Nayar, {\it The Review of Economics and
  Statistics\/} {\bf 101}, 60?75 (2018).

\bibitem{Inoue_Todo_2019}
H.~Inoue, Y.~Todo, {\it Nature Sustainability\/} {\bf 2}, 841?847 (2019).

\bibitem{noy2009macroeconomic}
I.~Noy, {\it Journal of Development economics\/} {\bf 88}, 221 (2009).

\bibitem{kellenberg2008does}
D.~K. Kellenberg, A.~M. Mobarak, {\it Journal of urban economics\/} {\bf 63},
  788 (2008).

\bibitem{maas2019facebook}
P.~Maas, {\it et~al.\/}, {\it Proceedings of the 16th International Conference
  on Information Systems for Crisis Response and Management (ISCRAM), Valencia,
  Spain. 2019\/} (2019).

\bibitem{sciencemobility}
C.~O. Buckee, {\it et~al.\/}, {\it Science\/}  (2020).

\bibitem{barabasi2016network}
A.-L. Barab{\'a}si, {\it et~al.\/}, {\it Network science\/} (Cambridge
  university press, 2016).

\bibitem{newman2018networks}
M.~Newman, {\it Networks\/} (Oxford university press, 2018).

\bibitem{latora2001efficient}
V.~Latora, M.~Marchiori, {\it Physical review letters\/} {\bf 87}, 198701
  (2001).

\bibitem{crucitti2006centrality}
P.~Crucitti, V.~Latora, S.~Porta, {\it Physical Review E\/} {\bf 73}, 036125
  (2006).

\bibitem{bullmore2009complex}
E.~Bullmore, O.~Sporns, {\it Nature reviews neuroscience\/} {\bf 10}, 186
  (2009).

\bibitem{deprivazione}
{Soluzioni per il Sistema Economico S.p.a (SOSE)}, Revisione della metodologia
  dei fabbisogni standard dei comuni,
  \url{http://www.mef.gov.it/ministero/commissioni/ctfs/documenti/Nota_revisione_metodologia_FS2017_SOSE_13_settembre_2016.pdf}
  (2016).

\bibitem{caranci_[italian_2010}
N.~Caranci, {\it et~al.\/}, {\it Epidemiologia E Prevenzione\/} {\bf 34}, 167
  (2010 Jul-Aug).

\bibitem{Koenker_Hallock_2001}
R.~Koenker, K.~F. Hallock, {\it Journal of Economic Perspectives\/} {\bf 15},
  143?156 (2001).

\end{thebibliography}

\section*{Acknowledgements}
\textbf{Funding:} A.S., A.F, G.B, F.PI, M.C. and W.Q. acknowledge the support from CNR P0000326 project AMOFI (Analysis and Models OF social medIa) and CNR-PNR National Project DFM.AD004.027 `` Crisis-Lab''.  \textbf{Authors contributions:} 
F.PI, AL.S, A.G, CM.V collected data; G.B, F.PI, M.C, A.G, A.S,  CM.V, F.PO analyzed data;
G.B, F.PI, M.C, A.F, W.Q, F.PA wrote the paper; W.Q, A.S, F.PA supervised the work.
 \textbf{Competing interests:} 
 authors have no competing interests, this should also be declared. 
 \textbf{Data and materials availability:} All data are made available for researcher to reproduce the result. Except for Facebook mobility data. This data was given in a previous agreement with Facebook.


\newpage
\section*{Figures}
\begin{figure}[!t]
    \centering
    \includegraphics[width=\linewidth]{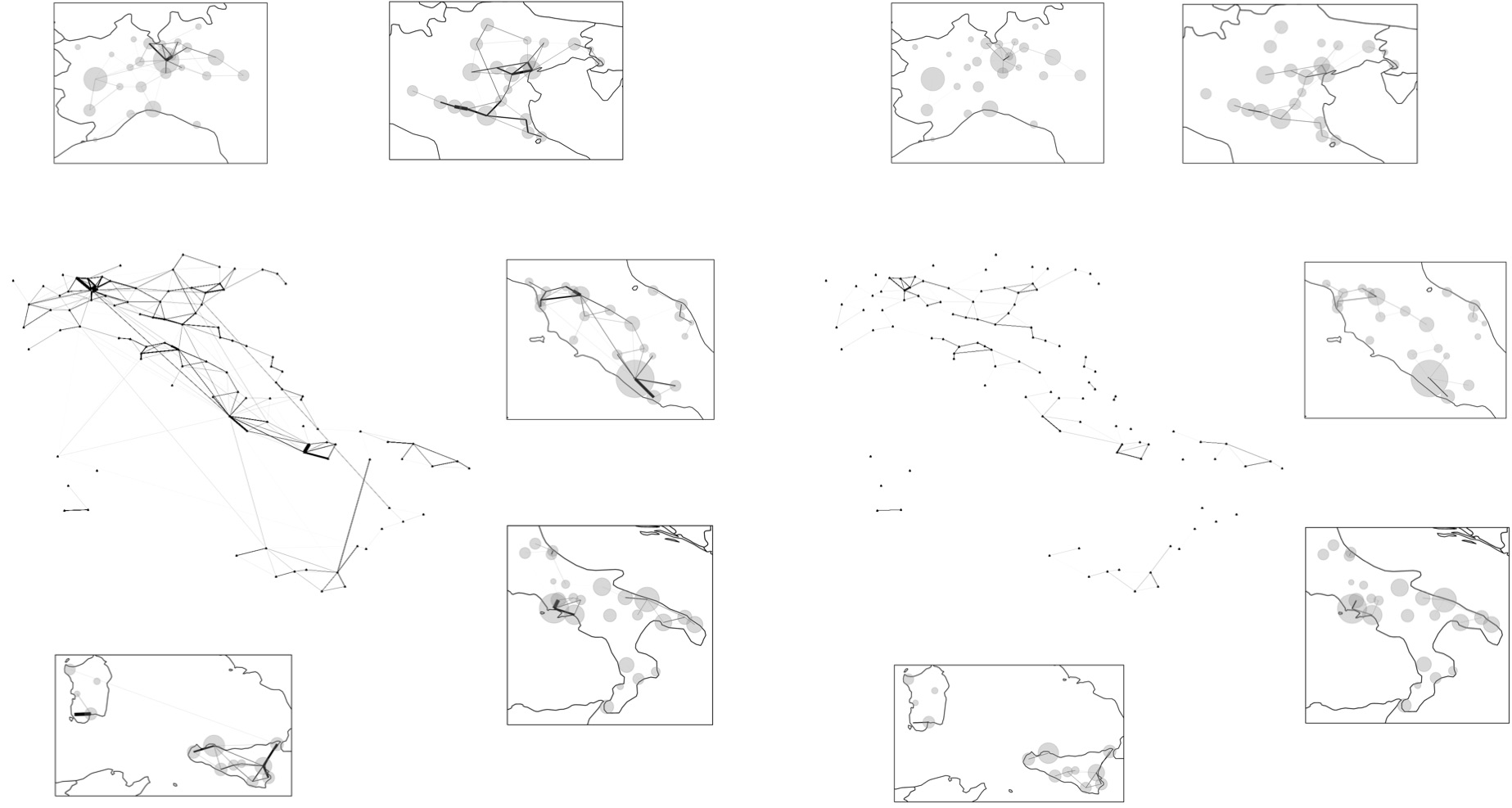}\\
    {\small\textbf{(A)} February 24th} \hspace{8cm} {\small\textbf{(B)} March 30th}
    \\
    \includegraphics[width=0.7\linewidth]{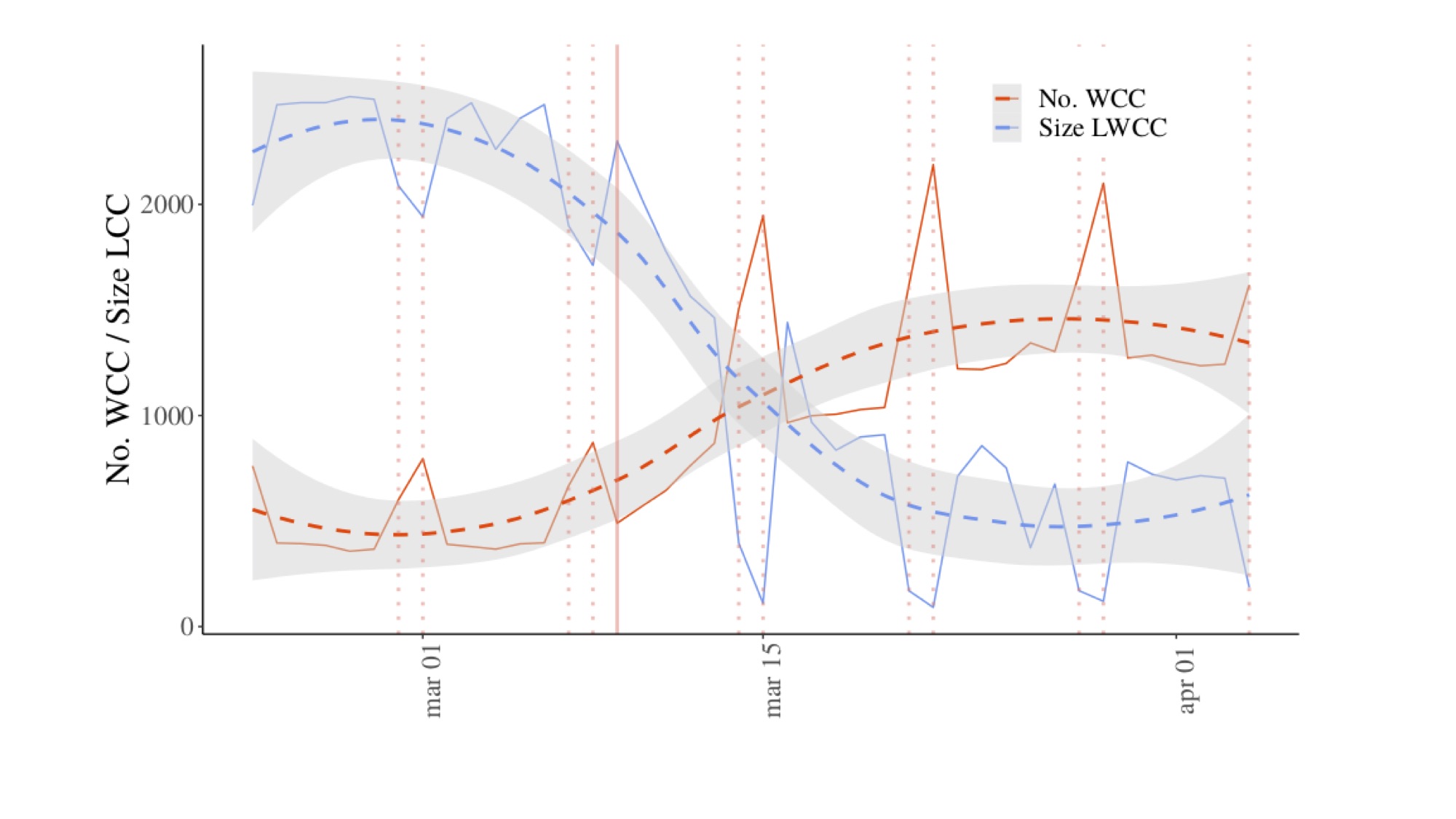} \\
    \textbf{(C)} Temporal evolution of network connectivity
    \caption{\textbf{Connectivity of the Italian mobility network during COVID-19 epidemic.}\\
    We provide a snapshot of the mobility network on two Mondays before and after national lockdown (9th March), respectively on February 24th (A) and March 30th (B). Nodes represent municipalities aggregated at province level, and they all have equal size, whereas thickness of edges is proportional to the weight. Inserts provide an outlook on different regions, where node size is instead proportional to the population of the province. Figure (C) represents the temporal evolution of the network connectivity in terms of number of weakly connected components (No. WCC, red) and size of the giant connected component (Size LWCC, blue), measured on daily snapshots of the mobility network since 23th February. To visualize trends, we show a LOESS regression (dotted line) with 95\% CI (shaded area), and highlight lockdown and week-days respectively with a solid and dotted vertical red lines. Trends are respectively significant increasing (M-K: $P\sim0$; K-T: $P\sim 0$ $R=0.64$; T-S: $R=30.52$) and decreasing (M-K: $P\sim0$; K-T: $P\sim 0$, $R=-0.67$; T-S: $R=-58.58$).
}
\end{figure}

\clearpage

\begin{figure}[!t]
    \centering
    \includegraphics[width=\linewidth]{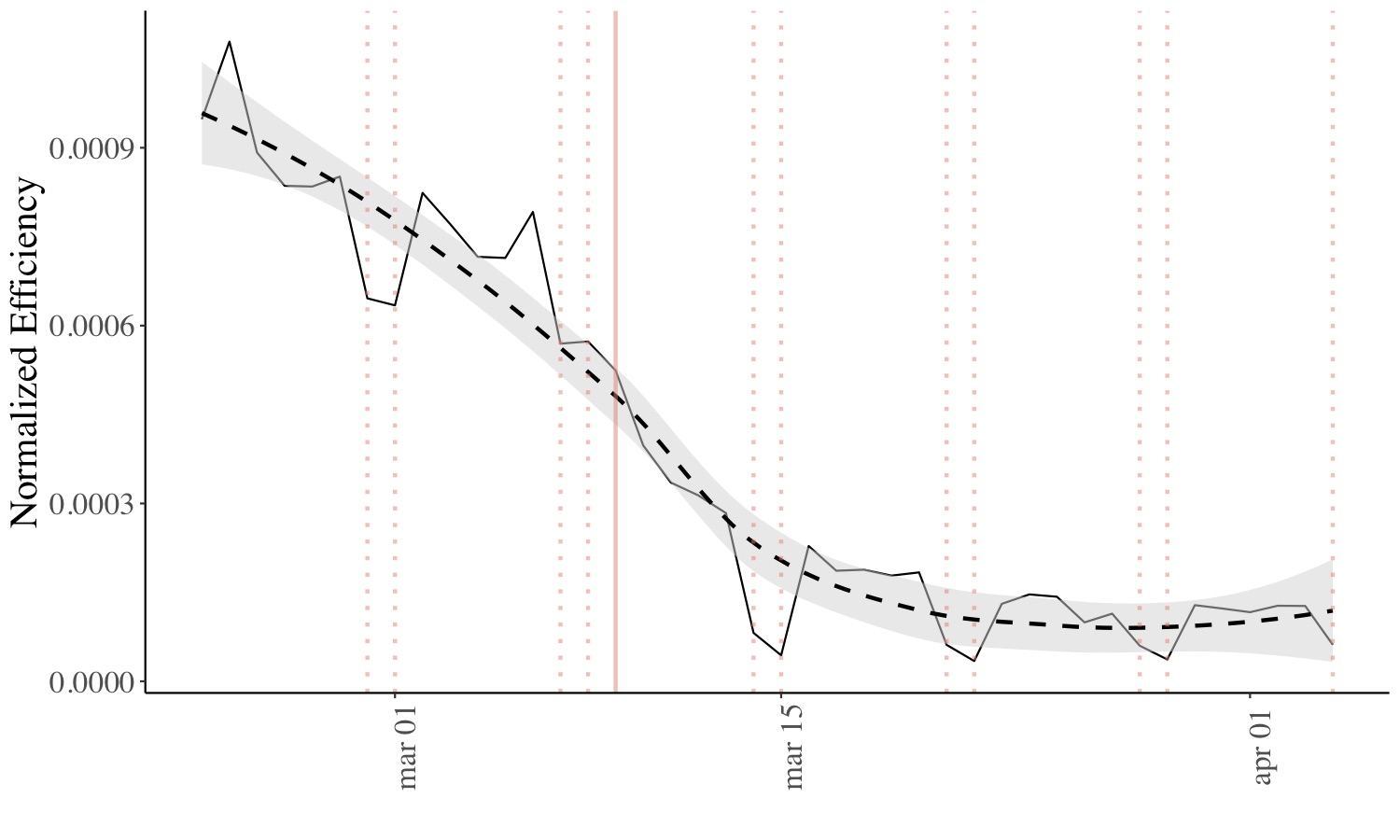}
    \caption{\textbf{Efficiency of the Italian mobility network during COVID-19 epidemic.}\\
    Temporal evolution of the global efficiency for the Italian mobility network from February 23rd to April 4th. Efficiency is computed according to \cite{latora2001efficient}. We use the reciprocal of weights to model distances between nodes. To visualize the trend, we show a LOESS regression (dotted line) with 95\% CI (shaded area), and highlight lockdown and week-days respectively with a solid and dotted vertical red lines. The trend is significantly decreasing (M-K: $P\sim0$; K-T: $P\sim 0$, $R=-0.75$; T-S: $R=-0.00003$).}
\end{figure}

\clearpage

\begin{figure}[!t]
    \centering
    \includegraphics[width=0.5\linewidth]{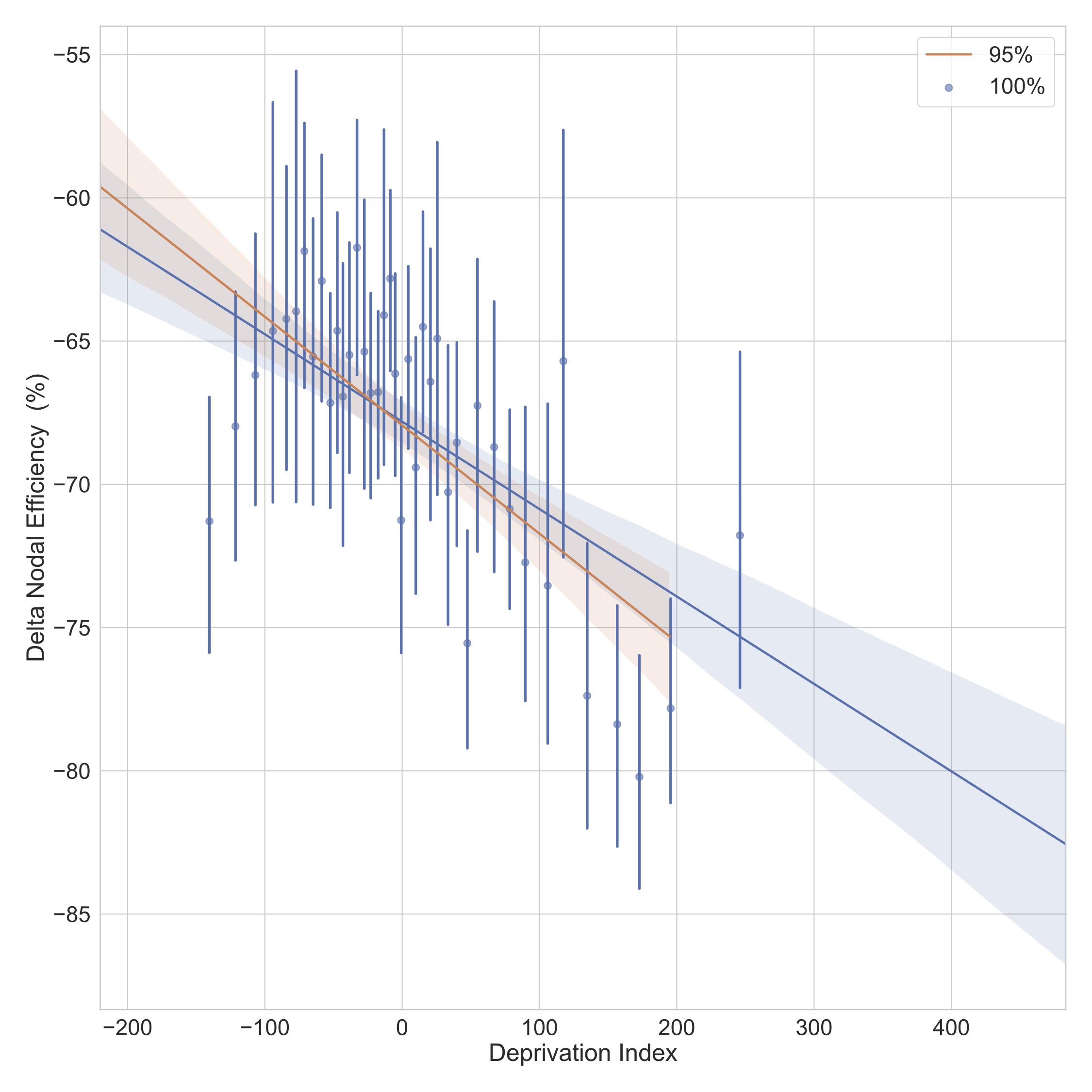}
    \\
    \includegraphics[width=0.5\linewidth]{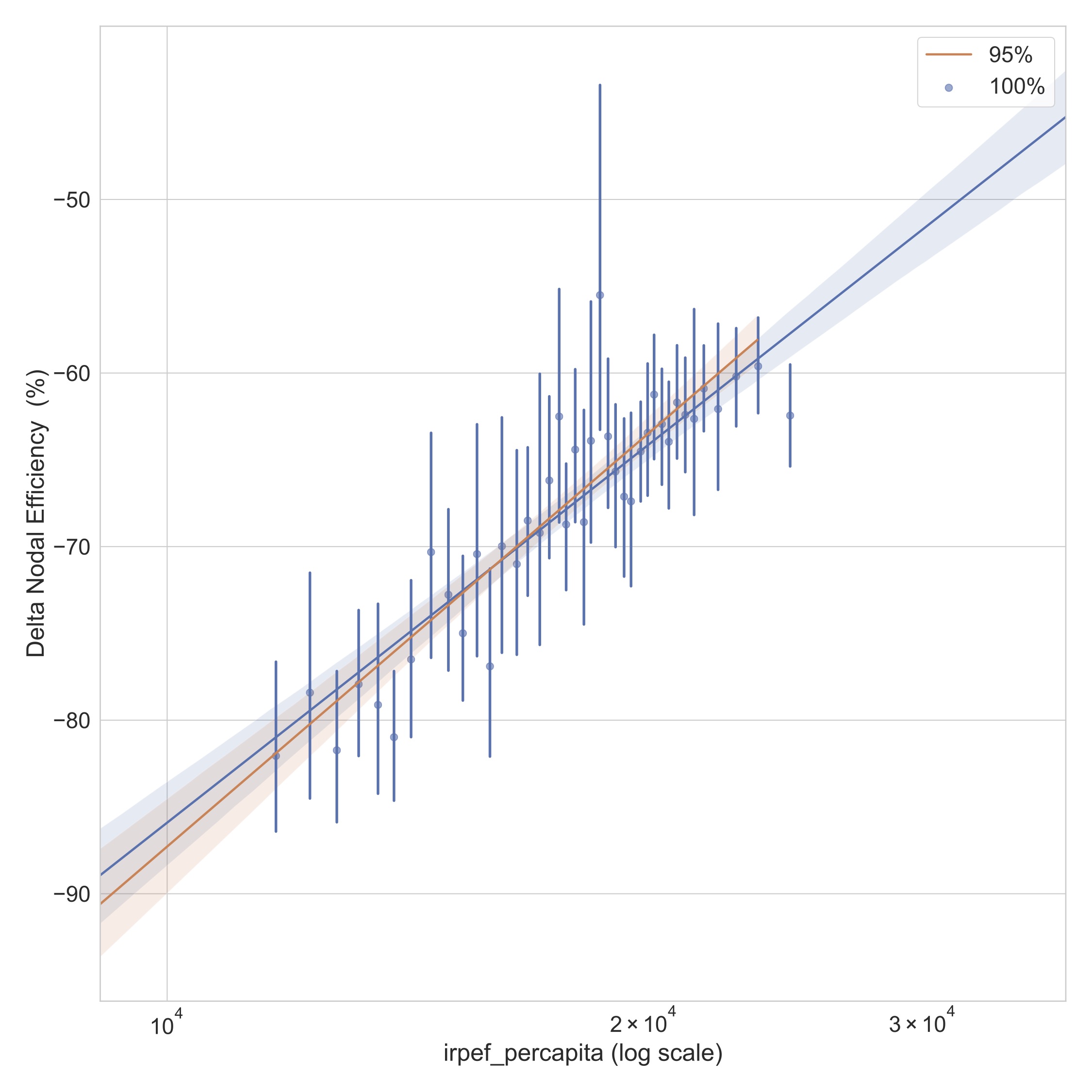} 
    \caption{\textbf{Correlation of mobility reduction and economic indexes.}\\
    Plot of the relative change in Nodal Efficiency, between pre- and post- lockdown values for each municipality, versus two different economic indexes, respectively Deprivation Index (top) and Irpef per Capita (bottom, x-axis in logarithmic scale). For each figure we provide two linear regression lines, one computed on the entire data (in blue) and one computed after filtering outliers w.r.t the economic index (in orange, using the 95-percentile as threshold). To ease the visualization, we draw the scatterplot by grouping points into 40 discrete bins, and showing mean and 95\% confidence interval of each bin. Correlations are respectively significantly negative
    (Pearson: $R=-0.153$, $P\sim0$; S-R: $R=-0.235$, $P\sim0$; T-S: $R=-0.064$; K-T: $R=-0.162$, $P\sim0$)  and significantly positive (Pearson: $R=0.263$, $P\sim0$; S-R: $R=0.404$, $P\sim0$; T-S: $R=0.001$; K-T: $R=0.273$, $P\sim0$).}\label{correlation}
\end{figure}

\clearpage


\begin{table}[ht!]
    \centering
    \footnotesize
    \resizebox{0.9\textwidth}{!}{\begin{tabular}{cccccccc}
\toprule
\hline
 q &   intercept & income pc & deprivation & fiscal capacity & inequality & real estate pc & (pseudo)R2 \\
\midrule
0.05   &  -0.8398*** &       0.2587*** &                      0.1686*** &  -0.1461*** &    -0.0344* &                  -0.1622*** &    0.05223 \\
 &    (0.0491) &        (0.0253) &                       (0.0276) &    (0.0286) &    (0.0204) &                    (0.0251) &            \\
0.1    &  -0.5089*** &       0.2871*** &                      0.1723*** &  -0.1280*** &    -0.0315* &                  -0.2539*** &    0.17578 \\
  &    (0.0456) &        (0.0260) &                       (0.0266) &    (0.0261) &    (0.0177) &                    (0.0232) &            \\
0.2    &  -0.2241*** &       0.2272*** &                      0.1272*** &  -0.0972*** &  -0.0410*** &                  -0.2907*** &    0.29896 \\
  &    (0.0317) &        (0.0187) &                       (0.0179) &    (0.0242) &    (0.0124) &                    (0.0206) &            \\
\midrule  
0.8    &   0.3770*** &       0.0788*** &                         0.0068 &  -0.0548*** &      0.0018 &                   0.0868*** &    0.14346 \\
  &    (0.0199) &        (0.0121) &                       (0.0117) &    (0.0123) &    (0.0094) &                    (0.0133) &            \\
0.9    &   0.8644*** &      -0.0962*** &                     -0.0868*** &  -0.3598*** &   0.2099*** &                   0.8759*** &    0.20012 \\
  &    (0.0523) &        (0.0347) &                       (0.0319) &    (0.0335) &    (0.0269) &                    (0.0304) &            \\
0.95   &   1.2128*** &      -0.2489*** &                      -0.1214** &  -0.3844*** &   0.3488*** &                   1.0334*** &    0.24347 \\
 &    (0.0761) &        (0.0542) &                       (0.0506) &    (0.0362) &    (0.0329) &                    (0.0345) &            \\
\midrule
OLS    &     0.1098* &        0.0654** &                        0.0557* &  -0.2045*** &   0.0737*** &                   0.1145*** &    0.09001 \\
    &    (0.0576) &        (0.0327) &                       (0.0328) &    (0.0404) &    (0.0239) &                    (0.0398) &            \\
\hline    
\bottomrule
\end{tabular}}

    \caption{\small Results for quantile regression of the relative difference of efficiency over time with respect to income per capita with multiple controls: social and financial distress in the municipality (deprivation and fiscal capacity), concentration of estates (real estate pc), income inequality and regional controls (extended model). Regression obtained with the Iterative Weighted Least Squares method on standardized variables. Standard errors reported in parenthesis calculated via bootstrap with 1000 iterations. Pseudo R2 obtained via McFadden's method. Only quantile 5-20 and 80-95 shown, full results available upon request. Bottom line shows OLS regression as reference. Number of observations: 2345. Not shown: coefficients of 19 regional controls.}
    \label{qreg_m2}
\end{table}

\clearpage
\renewcommand{\thesubfigure}{\Alph{subfigure}}

\begin{figure}
     \centering
     \begin{subfigure}[b]{0.475\textwidth}
         \centering
         \includegraphics[width=\textwidth]{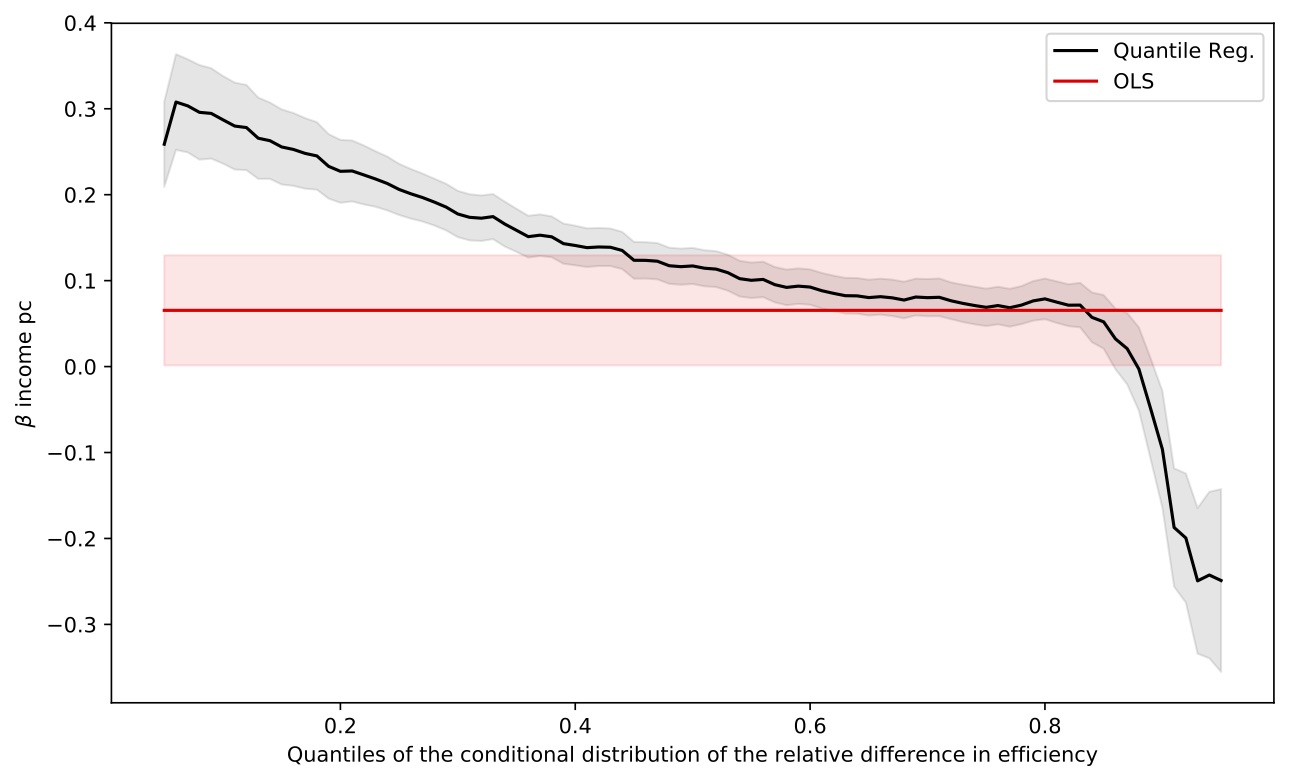}
         \caption{Income per capita}
         \label{fig:coeff_income}
     \end{subfigure} 
     \hfill
     \begin{subfigure}[b]{0.475\textwidth}
         \centering
         \includegraphics[width=\textwidth]{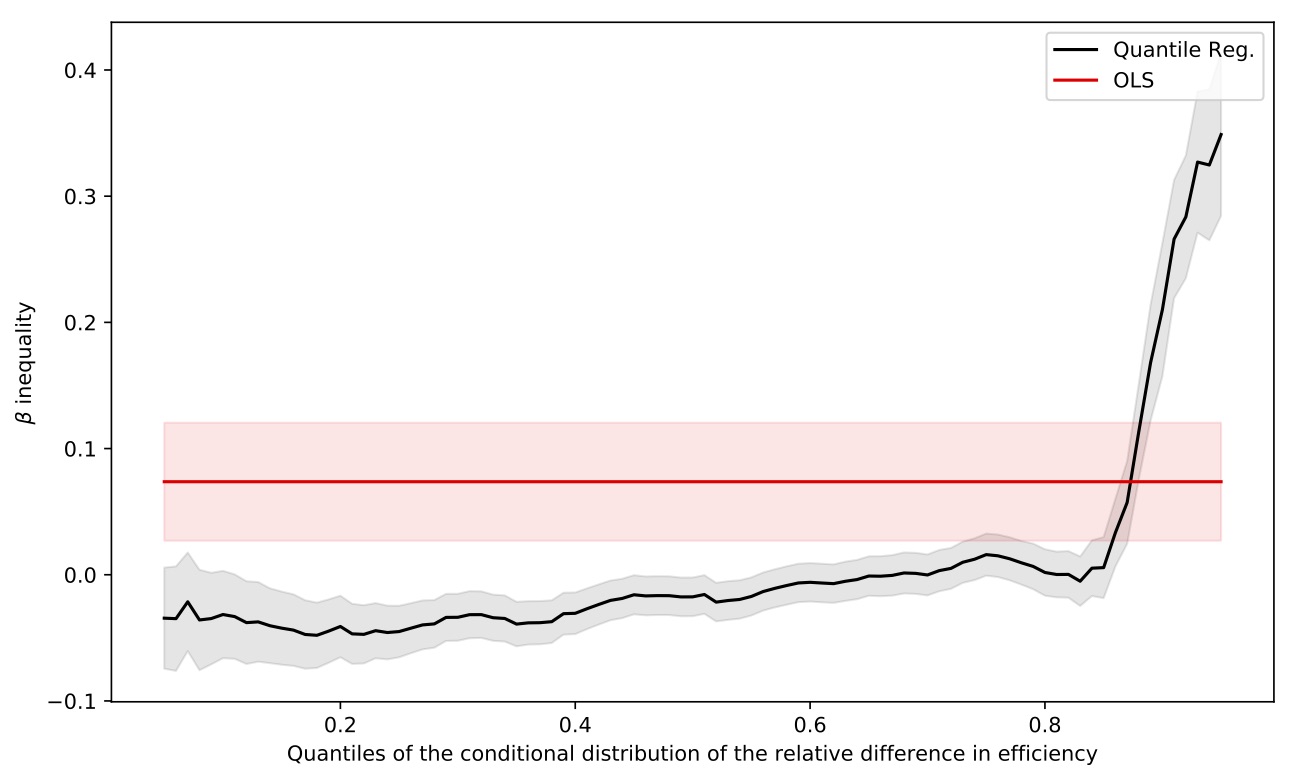}
         \caption{Inequality of income}
         \label{fig:coeff_ineq}
     \end{subfigure}
     \vskip\baselineskip
     \begin{subfigure}[b]{0.475\textwidth}
         \centering
         \includegraphics[width=\textwidth]{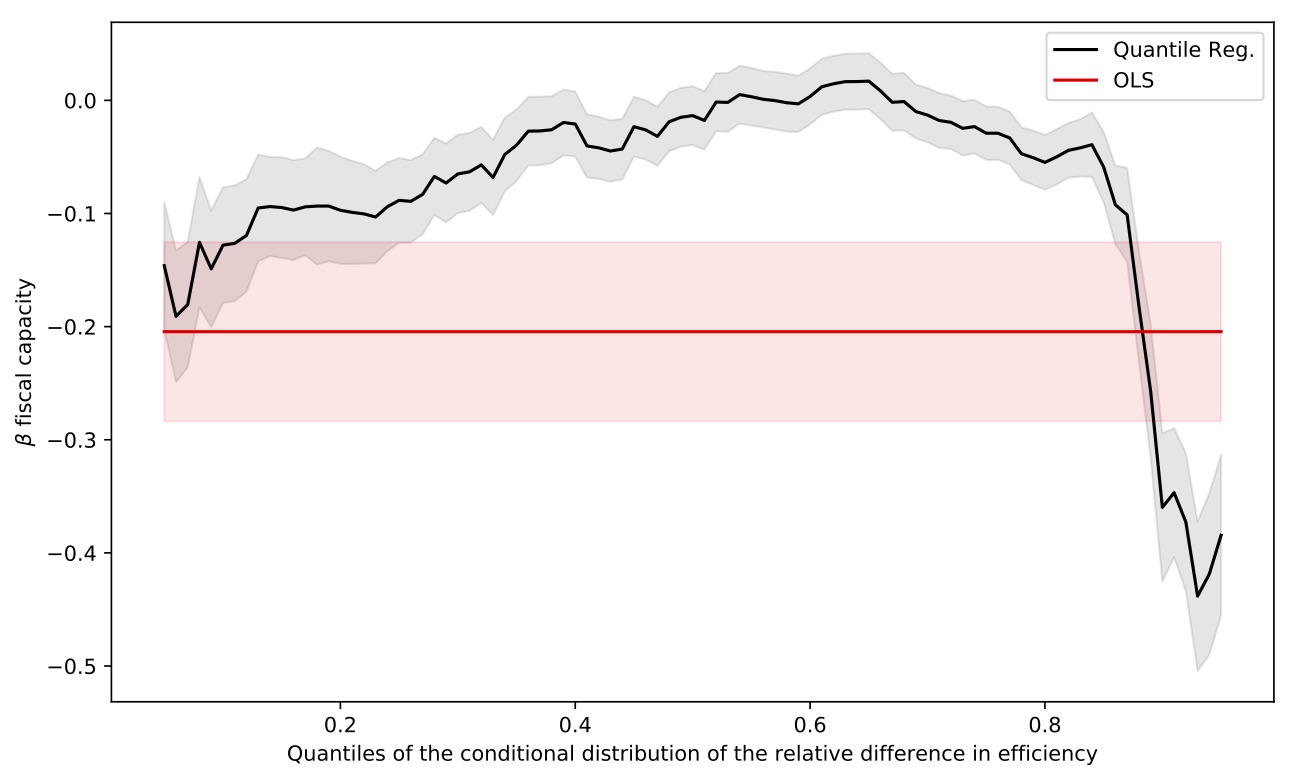}
         \caption{Fiscal capacity}
         \label{fig:coeff_fcap}
     \end{subfigure} 
     \hfill
     \begin{subfigure}[b]{0.475\textwidth}
         \centering
         \includegraphics[width=\textwidth]{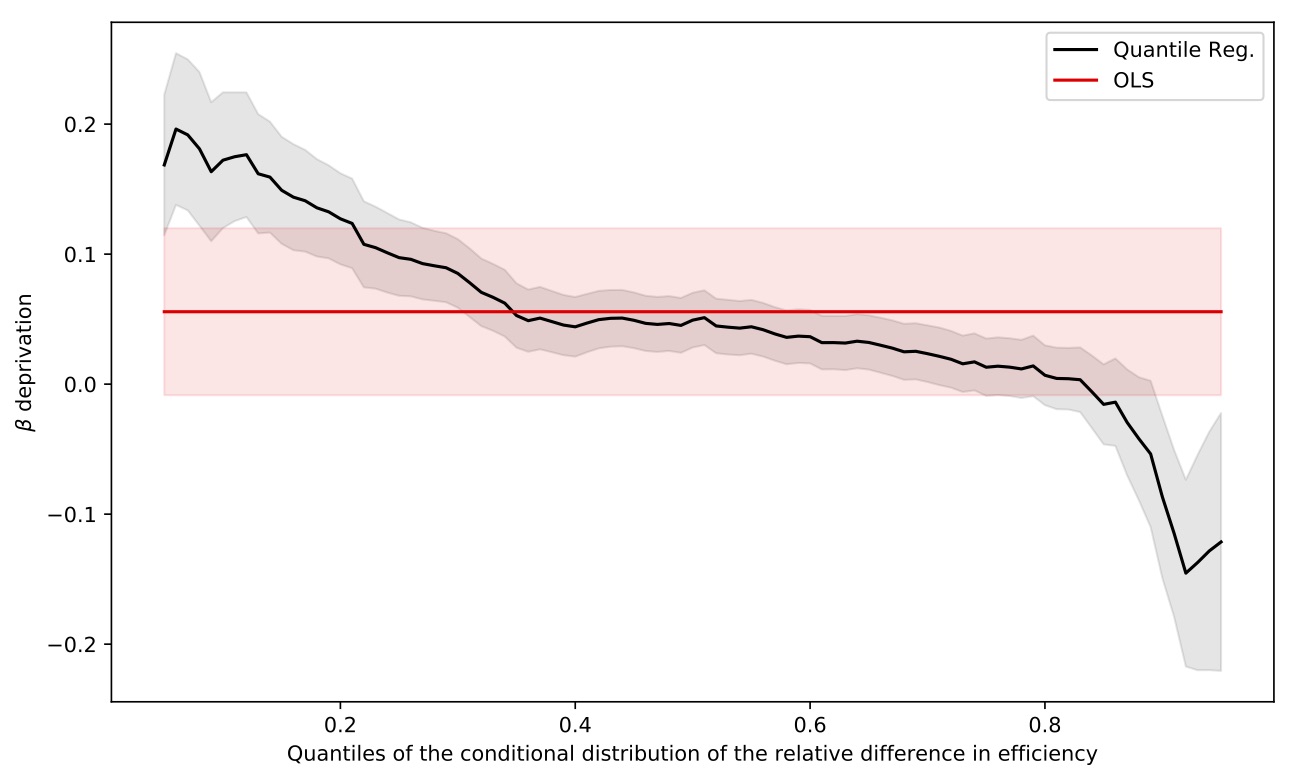}
         \caption{Index of deprivation}
         \label{fig:coeff_depriv}
     \end{subfigure}\\     
     
        \caption{\textbf{Regression coefficients by quantile for the extended model.}\\ 
        Plot of the regression coefficients for each percentile of the relative change in nodal efficiency over time for three of the main independent variables: Income per Capita (A), income Inequality (B), Fiscal capacity (C), Index of Deprivation (D). For each figure the coefficient of the quantile regression is plotted in black lines with bootstrapped confidence intervals at 95\%, as a reference the OLS regression of the same variables is plotted in red.  
        }\label{fig:three_coefficients}
\end{figure}

\restoregeometry

\end{document}